\documentclass[a4paper,11pt]{article}
\usepackage{jheppub}
\usepackage{array}
\usepackage{stmaryrd}
\usepackage[compat=1.1.0]{tikz-feynman}

\newcommand{\sq}[1]{[#1]}
\newcommand{\an}[1]{\langle#1\rangle}
\newcommand{\ket}[1]{|#1\rangle}
\newcommand{\res}[1]{\underset{#1}{\rm Res}}
\newcommand{\mo}{\mathcal{O}}
\newcommand{\mA}{\mathcal{A}}
\newcommand{\zb}{\bar{z}}
\newcommand{\D}{\Delta}
\newcommand{\e}{\epsilon}
\newcommand{\om}{\omega}
\newcommand{\bsym}[1]{\boldsymbol{#1}}
\newcommand{\tell}{\tilde{\ell}}

\def\mc{\mathcal}
\def\half{\frac{1}{2}}
\def\p{\partial}
\def\be{\begin{equation}}
\def\ee{\end{equation}}

\newcolumntype{L}{>{$}l<{$}}

\title{Supersymmetry and the Celestial Jacobi Identity}
\author[a,b]{Adam Ball,}
\author[a,c,d]{Marcus Spradlin,}
\author[a,e]{Akshay Yelleshpur Srikant}
\author[a,c]{and\\Anastasia Volovich}

\affiliation[a]{Department of Physics,
	Brown University,
	Providence, RI 02912, USA}
\affiliation[b]{Perimeter Institute for Theoretical Physics,
	Waterloo, ON N2L 2Y5, Canada}
\affiliation[c]{Department of Physics,
	Harvard University,
 	Cambridge, MA 02138, USA}
\affiliation[d]{Brown Theoretical Physics Center,
	Brown University,
	Providence, RI 02912, USA}
\affiliation[e]{Mathematical Institute, University of Oxford,
	Oxford, OX2 6GG, UK}

\emailAdd{aball1@perimeterinstitute.ca}
\emailAdd{marcus\_spradlin@brown.edu}
\emailAdd{Akshay.YelleshpurSrikant@maths.ox.ac.uk}
\emailAdd{anastasia\_volovich@brown.edu}

\abstract{In this paper we study the simplifying effects of supersymmetry on celestial OPEs at both tree and loop level. We find at tree level that theories with unbroken supersymmetry around a stable vacuum have celestial soft current algebras satisfying the Jacobi identity, and we show at one loop that celestial OPEs in these theories have no double poles.}

\begin{document}
\maketitle
	
\section{Introduction}
\label{sec:intro}

The celestial operator product expansion (OPE) has been the celestial holographer's work-horse since its inception in \cite{Fan:2019emx, Pate:2019lpp, Himwich:2021dau}. Of particular interest is its use in defining current algebras of celestial soft currents in Yang-Mills theory and Einstein gravity \cite{Guevara:2021abz}, which imply the existence of infinitely many symmetries constraining physical observables. In the gravitational case an appropriate redefinition of modes of positive-helicity soft gravitons, equivalent to a light transform, gives rise to the celebrated $w_{1+\infty}$-wedge current algebra \cite{Strominger:2021lvk}. However, it was shown in \cite{Mago:2021wje} that the analogous construction in a generic effective field theory (EFT) leads to a na\"ive current algebra that fails the Jacobi identity unless the coupling constants satisfy particular conditions. In practice, it is usually easier to diagnose this failure of Jacobi with the so-called double residue condition in momentum space \cite{Ren:2022sws},
\be \label{eq:dbc} 0 \stackrel{?}{=} \left( \res{z_2\shortrightarrow z_3} \, \res{z_1\shortrightarrow z_2} - \res{z_1\shortrightarrow z_3} \, \res{z_2\shortrightarrow z_3} + \res{z_2\shortrightarrow z_3} \, \res{z_1\shortrightarrow z_3} \right) {\cal A}_n \ee
where the $z_i$ refer to the following parametrization of (future-directed) massless momenta:
\begin{align}
    p_i = \omega_i \left(1+z_i \zb_i, z_i+\zb_i, z_i - \zb_i, 1-z_i \zb_i \right), \qquad i = 1, \dots, n.
\end{align}
This is a valid parametrization in split signature spacetime in which $z_i, \zb_i$ are real and independent coordinates. The ``holomorphic" OPE of the corresponding celestial operators is obtained in the limit $z_{12} \to 0$. This is also a limit in which (the left spinors of) the momenta become collinear.

The failure of the double residue condition on hard celestial amplitudes was traced to the presence of a branch cut in the multicollinear limit \cite{Ball:2023sdz}. This motivated the authors to propose that the OPE  in a generic tree-level EFT takes the form 
\begin{align}
    \label{eq:2bosonOPE}
    \mo_{\D_1}^{J_1}\left(z_1, \zb_1 \right) \mo_{\D_2}^{J_2}\left(z_2, \zb_2 \right) \, \sim & \,\, \frac{C_a}{z_{12}} \mo^{J_a}_{\D_a} + \frac{C_b}{z_{12}^{\D_b}} {\cal R}^{J_b}_{\D'_b}.
\end{align}
Here $\mo_{\D_i}^{J_i}$ are a pair of massless bosons with helicities $J_i$ ($i=1,2$), the $C_a$ terms arise from three-point interactions in the bulk EFT as described in \cite{Fan:2019emx,Pate:2019lpp} and the $C_b$ term represents less well-understood branch cut terms associated with multi-particle operators. Importantly, it was shown that such terms are absent at tree-level when the double residue condition is satisfied. In this paper, we demonstrate that tree-level OPEs in supersymmetric EFTs always satisfy the double residue condition implying a simpler OPE than their non-supersymmetric counterparts.\footnote{Supersymmetric OPEs and algebras for some supersymmetric cases have been computed in \cite{Ahn:2021erj, Ahn:2022oor, Drozdov:2023qoy}.}

The discussion so far has been at tree level, but it is important to ask also what happens at one loop. In pure Yang-Mills and Einstein gravity, the positive-helicity sectors are relative safe havens from corrections, and we can think of the number of minus-helicity legs in an amplitude as a rough measure of its complexity. In the positive-helicity sector of Einstein gravity, which at one loop is equivalent to self-dual gravity, amplitudes remain rational \cite{Bern:1998xc} and there are no corrections to the celestial OPE. Accordingly the $w_{1+\infty}$-wedge symmetry is preserved exactly \cite{Ball:2021tmb}. There is a closely analogous story for self-dual Yang-Mills and the positive-helicity sector of Yang-Mills \cite{ Bern:1994zx, Bern:1991aq, Bern:1993qk, Mahlon:1993si, Bern:1994ju}. Moving on to single-minus amplitudes or all-plus form factors, amplitudes are still rational \cite{Alston:2015gea, Mahlon:1993si} but the celestial OPE receives corrections in the form of double pole terms. The resulting na\"ive current algebra fails Jacobi \cite{Costello:2022upu, Bittleston:2022jeq}. The authors of \cite{Costello:2022upu} showed that Jacobi can be restored by the addition of a certain scalar with a quartic kinetic term. Moreover \cite{Costello:2022upu} also pointed out that carefully chosen matter content would achieve the same while \cite{Bittleston:2022jeq, Costello:2023vyy, Bittleston:2023bzp} noted that such OPEs are associative in self-dual supersymmetric theories. Attempts to venture even further from the self-dual sector, for example by studying maximally helicity violating (MHV) Yang-Mills amplitudes (which have two minus-helicity legs), have met with seemingly less well-behaved celestial OPEs that include both double poles and log terms \cite{Gonzalez:2020tpi, Bhardwaj:2022anh}. Other related aspects of loop-level celestial amplitudes have been studied in~\cite{Banerjee:2017jeg,Albayrak:2020saa,Gonzalez:2021dxw,Nastase:2021izh}.

Upon including loop corrections, the OPE (\ref{eq:2bosonOPE}) is generically modified to 
\begin{align}
    \mo_{\D_1}^{J_1}\left(z_1, \zb_1 \right) \mo_{\D_2}^{J_2}\left(z_2, \zb_2 \right) \, \sim & \,\, \frac{C_a}{z_{12}} \mo^{J_a}_{\D_a} + \frac{C_b}{z_{12}} \mo^{J_b}_{\D_b} +\nonumber \\
    & \frac{C_c}{z_{12}^2} \mo_{\D_c}^{J_c} + \frac{C_d}{z_{12}^{\D_d}} {\cal R}^{J_d}_{\D'_d} \\
    &\nonumber + \text{non-factorizing}.
\end{align}
The $C_b$ and $C_c$ terms are the single pole and double pole contributions arising from massless loops. The non-factorizing contributions involve logarithms of $z_{12}$ as well as derivatives with respect to the conformal dimensions. The explicit form of this correction for gluons can be found in \cite{Gonzalez:2020tpi,Bhardwaj:2022anh}. We will show that $C_c = 0$ for all bosonic OPEs in certain supersymmetric theories. While we cannot apply the double residue condition at one loop due to the non-analytic structure of the OPE, the vanishing of double poles is a concrete way in which CCFTs dual to SUSY theories are simpler than their non-SUSY counterparts, making them ideal candidates to further probe the effect of loop corrections. Indeed the power of SUSY has been well exploited and has led to enormous progress in the computation of multi-loop scattering amplitudes, and it seems natural that this should extend to Celestial Holography. Loop corrections also affect the soft theorems and thereby the corresponding Ward identities. In particular, the subleading soft graviton theorem, which is equivalent to the superrotation Ward identity, receives loop corrections from the IR divergent part of one-loop amplitudes as well as the finite part in generic theories of gravity \cite{Bern:2014oka, He:2014bga}. The finite corrections are derived from the single-minus amplitudes and thus vanish in supersymmetric theories, implying that the superrotation Ward identities are corrected only by the universal IR divergent terms. In some cases, it might be possible to incorporate supersymmetry naturally into the structure of the celestial CFT~\cite{Taylor:2023bzj}. 

In the rest of this paper, we will analyze the constraints imposed by SUSY on celestial OPEs. We will restrict to $\mc{N}=1$ SUSY and to the analysis of bosonic OPEs. After setting notation and recapitulating some basic facts about $\mc{N}=1$ SUSY theories in Section \ref{sec:prelims}, we prove in Section \ref{sec:treeOPE} that Jacobi holds at tree level for $\mc{N}=1$ SUGRA and all $\mc{N}\ge 1$ global SUSY EFTs.\footnote{We use ``global" to emphasize the absence of local SUSY, i.e. the absence of a graviton multiplet.} We then move on to describe the collinear structure of one-loop amplitudes using the simple example of pure Yang-Mills in Section \ref{sec:1loopfact}. Finally we present the cancellation of double poles at one loop in Section \ref{sec:dpcancel}.

\section{Preliminaries}
\label{sec:prelims}

In this work, we will focus on collinear singularities of scattering amplitudes in split signature or Kleinan spacetime and the corresponding celestial OPEs. Collinear particles are necessarily massless, and massive interactions can be integrated out, so for our purposes we restrict attention to massless theories. We find it convenient to adopt the following conventions. Coupling constants will be indicated explicitly as opposed to stripping them off as is commonplace in the literature on scattering amplitudes. Amplitudes involving coloured particles will implicitly be colour-ordered unless specified and colour indices will always be suppressed. Finally, we assume a perturbative expansion of amplitudes based on the number of loops and write
\begin{align}
    \mA_n \left(1^{J_1}, 2^{J_2}, 3^{J_3}, \dots \right) = \sum_{L=0}^{\infty} \mA_n^{(L)} \left(1^{J_1}, 2^{J_2}, 3^{J_3}, \dots \right).
\end{align}
Here $\mA_n^{(L)}$ represents the contribution from all $L$-loop Feynman diagrams, which we will refer to as the $L$-loop amplitude with $\mA_n^{(0)}$ being the tree-level contribution.

\subsection{Three-point amplitudes}

We will often make use of 3-point amplitudes and list them here for convenience. When all three particles are massless, these are given by
\begin{equation}
\label{eq:threepoints}
    \mA_3(1^{J_1},2^{J_2},3^{J_3}) = \left\{
    \begin{aligned}
        & \kappa_{J_1,J_2,J_3} \,\,\lbrack 12 \rbrack^{J_1+J_2-J_3} \lbrack 23 \rbrack^{J_2+J_3-J_1} \lbrack 31 \rbrack^{J_3+J_1-J_2} && \text{if } J_1+J_2+J_3 > 0, \\
        & \kappa_{J_1,J_2,J_3} \,\, \langle 12 \rangle^{J_3-J_1-J_2} \langle 23 \rangle^{J_1-J_2-J_3} \langle 31 \rangle^{J_2-J_1-J_3} && \text{if } J_1+J_2+J_3 < 0.
    \end{aligned}
    \right.
\end{equation}
where $\an{ij}$ and $\sq{ij}$ are the usual angle and square brackets of the spinor helicity variables associated with null momenta and $\kappa_{J_1, J_2, J_3}$ are the 3-point couplings of the theory with mass dimension $\left[\kappa_{J_1, J_2, J_3}\right] = 1 - \left|J_1+J_2+J_3\right|$. We ignore 3-point amplitudes with $J_1+J_2+J_3 = 0$ (unless all three $J_i=0$) since they do not lead to consistent 4-point amplitudes \cite{McGady:2013sga}. 

\subsection{Parametrizing the collinear limit}
\label{sec:collinearpar}

For collinear configurations of three particles with momenta labelled by $p_1, p_2$ and $-P$, momentum conservation and the on-shell condition imply
\begin{equation}
    \label{eq:collinearconfig}
    p_1 + p_2 - P = 0, \qquad p_1^2 = p_2^2 = P^2 = 0 \implies p_1 \cdot p_2 = \an{12} \sq{21} = 0.
\end{equation}
In split signature, $\an{12}$ and $\sq{12}$ are real and independent. In this paper, we will be interested in the ``holomorphic'' collinear limit in which $\an{12} \to 0$ with $\sq{12}$ held fixed. In this limit, we have
\begin{equation}
    \label{eq:collinearspinorrels}
    \frac{1}{\sq{2P}}\lambda_1 = \frac{1}{\sq{P1}}\lambda_2 =\frac{1}{ \sq{12}} \lambda_P.
\end{equation}
In order to make the transition to celestial amplitudes, we will adopt the following parametrization for outgoing massless momenta (valid in $(+,-,+,-)$ signature)
\begin{align}
    p_i = \omega_i \left(1+z_i \zb_i, z_i+\zb_i, z_i - \zb_i, 1-z_i \zb_i \right), \qquad i = 1, \dots, n.
\end{align}
The corresponding spinor helicity variables are
\begin{align}
    \label{eq:shvars}
    \lambda_i = \sqrt{2 \omega_i} \e_i\begin{pmatrix}
        1\\ z_i
    \end{pmatrix}, \qquad \tilde{\lambda}_i = \sqrt{2 \omega_i}\begin{pmatrix}
        1\\ \zb_i
    \end{pmatrix}, \qquad i = 1, \dots, n.
\end{align}
Finally, a convenient way to describe the collinear limit in this parametrization is
\begin{align}
    \label{eq:collinearlimitpar}
    \om_1 = t\, \om_P, \qquad \om_2 = (1-t) \, \om_P, \qquad z_1 = z_2 = z_P, \qquad \zb_P = t \zb_1 + (1-t) \zb_2.
\end{align}
We will always consider the particles $1$ and $2$ to be outgoing.

\subsection{\texorpdfstring{An $\mathcal{N}=1$ SUSY primer}{An N=1 SUSY primer}}
\label{sec:SUSYprimer}

In theories with $\mc{N}=1$ SUSY we can simultaneously diagonalize the momenta and one of the supercharges. States in such theories are thus labelled by the eigenvalues of a supercharge in addition to the momenta. In each massless multiplet there is a ``top" state of highest helicity and a ``bottom" state of one-half lower helicity. For example in the graviton multiplet $\{ \ket{+2}, \ket{+\frac{3}{2}} \}$ the top state is $\ket{+2}$, and in the multiplet $\{ \ket{-\half}, \ket{-1} \}$ the top state is $\ket{-\half}$.

We use the notation $\mA_n \left(\left\lbrace t_1^{J_1}, \dots, t_k^{J_k}\right\rbrace, \left\lbrace b_{k+1}^{J_{k+1}}, \dots, b_n^{J_n}\right\rbrace \right)$ for the $n$-point amplitude with top legs $t_i^{J_i}$ and bottom legs $b_{i}^{J_{i}}$. The superscripts $J_i$ on specify the helicities, and we suppress other labels. These theories are heavily constrained by supersymmetric Ward identities (SWIs) \cite{Grisaru:1977px, Grisaru:1976vm}. The SWIs say that amplitudes whose external legs are all top or all-but-one top must vanish, and similarly for bottom states. At four points this means nonzero amplitudes must have precisely two top and two bottom legs. At three points the SWIs partially break down, but it is still true that 3-point amplitudes with all-top or all-bottom legs vanish. These identities can be summarized as
\begin{align}
    &\mA_n \left(\left\lbrace t_1^{J_1} \dots, t_n^{J_n} \right\rbrace \right) = 0 \qquad n \geq 3, \\
    &\mA_n \left(\left\lbrace t_1^{J_1} \dots, t_{n-1}^{J_{n-1}} \right\rbrace, \left\lbrace b_n^{J_n} \right\rbrace \right) = 0 \qquad n \geq 4,
\end{align}
along with CPT conjugate relations which exchange $t_i^{J_i} \leftrightarrow b_i^{-J_i}$. These are non-perturbative statements and hold at all loop orders. In this paper, we will consider theories with arbitrary numbers of $\mc{N}=1$ SUSY multiplets of massless particles drawn from the following set:
\begin{equation}
\label{eq:multiplets}
    \big\{\ket{2}, \ket{\frac{3}{2}}\big\}, \ \big\{\ket{1}, \ket{\frac{1}{2}}\big\}, \ \big\{\ket{\frac{1}{2}}, 0\big\} \quad + \quad \text{CPT conjugates}.
\end{equation}
Note that positive-helicity bosonic states are at the top of their multiplets --- an observation that will prove useful in Section \ref{sec:treeOPE}. We have excluded the gravitino multiplet $\{\ket{\frac{3}{2}}, \ket{1}\}$, which arises only in extended SUGRA, because its graviphoton state $\ket{1}$ is at the bottom of its multiplet yet has positive helicity. This obstructs our argument for $\mc{N}\ge 2$ SUGRA, but nonetheless it is plausible that the large amount of supersymmetry would still protect the Jacobi identity. We leave the resolution of this question to future work. We make no assumptions on the gauge groups of the particles other than those required by SUSY.

\section{Tree-level supersymmetric EFTs}
\label{sec:treeOPE}

In this section we show that $\mc{N}=1$ SUGRA and $\mc{N}\ge 1$ global SUSY EFTs around stable vacua give rise to (bosonic) celestial soft currents satisfying the Jacobi identity at tree level.\footnote{Note that vacuum stability precludes 3-point couplings between massless scalars, since they would imply that the vacuum is merely a stationary point of the potential, as opposed to a local minimum.} In practice, it is usually easiest to work with momentum space amplitudes, and \cite{Ren:2022sws} showed that satisfaction of Jacobi implies the vanishing of the holomorphic all line shift constructible part of the amplitude. The converse was shown by \cite{Ball:2022bgg}, along with the observation that the whole question boils down to the angle bracket weight $-1$ parts of 4-point momentum space amplitudes, where the angle bracket weight of a term is defined by writing the amplitude in spinor-helicity form and counting the number of angle brackets in the numerator minus the number in the denominator.\footnote{The fact that all line shift constructibility requires the angle bracket weight of an amplitude to be negative was proved in \cite{Cohen:2010mi}.} For example the 4-point amplitude in $\phi^3$ theory is simply $\frac{1}{\langle 12 \rangle [12]} + \frac{1}{\langle 13 \rangle [13]} + \frac{1}{\langle 14 \rangle [14]}$, and all three terms have angle bracket weight $-1$, so the na\"ive celestial currents in $\phi^3$ theory do not satisfy the Jacobi identity.

We want to know the angle bracket weight $-1$ parts of our 4-point amplitudes, which is the same as the all line shift constructible part. A 3-point amplitude with positive helicity sum, $\sum_{i=1}^3 h_i > 0$, is a monomial of square bracket spinor products and therefore has angle bracket weight zero. A 3-point amplitude with negative helicity sum is a monomial of angle bracket spinor products, and therefore has positive angle bracket weight. 3-point amplitudes with helicity sum zero are believed to only be consistent for 3-scalar interactions, which are absent due to our assumption of vacuum stability. In the all-line shift construction one sews together two 3-point amplitudes with a massless propagator to get a 4-point amplitude. The propagator has angle bracket weight $-1$, and to maintain that weight for the full 4-point amplitude we must use only 3-point amplitudes with positive helicity sum in our construction.

Let us further refine the types of 3-point amplitudes that can contribute to a 4-point amplitude with nonzero angle bracket weight $-1$ part. In addition to having positive helicity sum, the SWIs also imply that they cannot be all-top. In order to sew together legs from two 3-point amplitudes, the legs must be CPT conjugates. This means that one is top and one is bottom. The other legs will become the external legs of the 4-point amplitude, of which precisely two must be top due to SWIs, so overall we see that precisely three of the six legs of our two 3-point amplitudes must be top. This means that one of the 3-point amplitudes must have a single top leg. We have finally arrived at our key condition: we need at least one 3-point amplitude with a positive helicity sum but only one top leg. The only such massless 3-point amplitudes are of the following types:
\be \mA_{++, \, \bar\phi, \, \bar\phi}, \qquad \mA_{++, \, -, \, \bar\phi}, \qquad \mA_{+, \, \bar\phi, \, \bar\phi}. \ee
Here we use $++$ to denote a positive-helicity graviton, $+$ to denote a positive-helicity gluon/photon, $\phi$ to denote a scalar that is at the top of its multiplet, and $\bar\phi$ to denote a scalar at the bottom of its multiplet. There may be many species of gluon/photon/scalar. We have used the fact that positive-helicity gluons/photons in $\mc{N}=1$ SUGRA are necessarily at the tops of their multiplets, and similarly in $\mc{N}\ge 1$ global SUSY theories. To rule out these amplitudes we examine explicit Lagrangians.

The most general $\mc{N}=1$ SUGRA Lagrangian is given in (G.2) of \cite{Wess:1992cp}, and one can read off 3-point couplings to find that
\be \label{eq:vanishingamps} 0 = \mA_{++, \bar\phi, \bar\phi} = \mA_{++, -, \bar\phi} = \mA_{+, \bar\phi, \bar\phi}. \ee
Therefore all 4-point amplitudes have vanishing angle bracket weight $-1$ part, and consequently Jacobi is satisfied in $\mc{N}=1$ SUGRA around a stable vacuum. In global SUSY theories there are no gravitons, so our only worry is the amplitude $\mA_{+,\bar\phi,\bar\phi}$. It would need to come from an operator like $A_\mu \bar\phi \p^\mu \bar\phi$ (possibly with different species of $\bar\phi$), which has mass dimension four and is therefore renormalizable. The most general renormalizable $\mc{N}=1$ global SUSY Lagrangian is given in (10.53) and (11.4a) of \cite{Sohnius:1985qm}, and we can see that such a term is absent. Note this applies also to theories with $\mc{N} > 1$ global SUSY, since they are special cases of $\mc{N}=1$. Therefore Jacobi is satisfied in $\mc{N}\ge 1$ global SUSY EFTs around a stable vacuum.

\section{Review of collinear factorization at one loop}
\label{sec:1loopfact}

OPEs in CCFT are inherited from the collinear structure of scattering amplitudes. In this section, we will review the general structure of collinear limits of one-loop amplitudes which has been well-studied in gauge theory and gravity \cite{Mangano:1990by, Bern:1994zx, Bern:1995ix, Bern:1998sc, Kosower:1999rx, Bern:1999ry, Kosower:1999xi, Akhoury:2011kq}. With pedagogy in mind, we will work with the simple example of pure Yang-Mills theory. The generalizations to other theories including gravity are straightforward. 

The 3-point amplitudes of this theory can be read off from (\ref{eq:threepoints}) to be
\begin{align}
    \label{eq:YM3pt}
    &\mA_3^{(0)} \left(1^{+}, 2^{+}, 3^{-} \right) = \kappa_{1,1,-1}\frac{\sq{12}^3}{\sq{23}\sq{31}}, \qquad \mA_3^{(0)} \left(1^{-}, 2^{-}, 3^{+} \right) = \kappa_{1,1,-1}\frac{\an{12}^3}{\an{23}\an{31}},
\end{align}
In writing the above equations, we have suppressed the colour factors and set the two 3-gluon couplings $\kappa_{1,-1,-1} = \kappa_{1,1,-1}$ equal to preserve parity. Note that
\begin{equation}
    \label{eq:minYMzero}
    \mA_3^{(0)} \left(1^{+}, 2^{+}, 3^{+} \right) = \mA_3^{(0)} \left(1^{-}, 2^{-}, 3^{-} \right) = 0
\end{equation}
since $\left[\kappa_{1,1,1}\right] = \left[\kappa_{-1,-1,-1}\right] = -2$ and their presence would be indicative of higher dimension operators. In the collinear limit, we have at tree level
\begin{equation}
    \mA^{(0)}_n \left(1^{+}, 2^{+}, 3, \dots n \right) = \sum_{J = \pm 1} \text{Split}^{(0)} \left(1^{+}, 2^{+}; -P^{-J} \right)\mA^{(0)}_{n-1} \left(P^{J}, 3, \dots n \right).
\end{equation}
The splitting function can be easily determined from 3-point amplitudes as a consequence of unitarity which dictates that the residue on the two-particle pole $\left(p_1 + p_2 \right)^2 \to 0$ is
\begin{equation}
\label{eq:treefact}
    \mA^{(0)}_n \left(1^{+}, 2^{+}, 3, \dots n \right) \xrightarrow[]{\left(p_1 + p_2 \right)^2 \to 0} \sum_{J=\pm 1} \mA_3^{(0)} \left(1^{+}, 2^{+}, -P^{-J} \right) \, \frac{1}{\an{12}\sq{21}} \mA^{(0)}_{n-1} \left(P^{J}, 3, \dots n \right).
\end{equation}
The splitting functions can thus be read off as
\begin{align}
    \label{eq:ppmtree}&\text{Split}^{(0)} \left(1^{+}, 2^{+}; -P^{-} \right) = \mA_3^{(0)} \left(1^{+}, 2^{+}, -P^{-1} \right) \, \frac{1}{\an{12}\sq{21}} = \frac{\kappa_{1,1,-1}}{\an{12}}\frac{\sq{12}^2}{\sq{2P}\sq{P1}},\\
    \label{eq:ppptree}&\text{Split}^{(0)} \left(1^{+}, 2^{+}; -P^{+} \right) = \mA_3^{(0)} \left(1^{+}, 2^{+}, -P^{+1} \right) \, \frac{1}{\an{12}\sq{21}} = 0.
\end{align}
In arriving at the above results we have used (\ref{eq:YM3pt}) and (\ref{eq:minYMzero}). It is easy to check that after using the parametrization in (\ref{eq:collinearlimitpar}), the splitting functions reduce to the standard ones \cite{Mangano:1990by}.

The collinear behaviour at one loop is
\begin{multline}
     \mA^{(1)}_n \left(1^{+}, 2^{+}, 3, \dots n \right) = \sum_{J = \pm 1} \text{Split}^{(0)} \left(1^{+}, 2^{+}; -P^{-J} \right)\mA^{(1)}_{n-1} \left(P^{J}, 3, \dots n \right) \\
     +\sum_{J = \pm 1} \text{Split}^{(1)} \left(1^{+}, 2^{+}; -P^{-J} \right) \mA^{(0)}_{n-1} \left(P^{J}, 3, \dots n \right).
\end{multline}

We now focus on the the corrections from a gluon loop. There are two distinct contributions --- factorizing and non-factorizing --- each associated with its own splitting function,
\begin{equation}
\label{eq:1loopsplit}
    \text{Split}^{(1)} \left(1^{+}, 2^{+}; -P^{-J} \right) = \text{Split}_{\text{fact}}^{(1)} \left(1^{+}, 2^{+}; -P^{-J} \right) + \text{Split}_{\text{non-fact}}^{(1)} \left(1^{+}, 2^{+}; -P^{-J} \right).
\end{equation}
The factorizing contributions arise from gluon loop corrections to the 3-point amplitude (with one leg initially off-shell). We present the details of this computation in Appendix \ref{app:3pt-1loop-gl}, merely displaying the result that leads to the double pole below
\begin{align}
   \label{eq:1loopgl}
    &\mA_3^{(1)} \left(1^+,2^+,-P^+\right) \xrightarrow[]{\an{12} \to 0} \frac{N_g \, \kappa_{1,1,-1}^3}{16\pi^2}\frac{\sq{P1}\sq{2P}}{\an{12}} + \mathcal{O} \left(\an{12}^0\right).
\end{align}
The one-loop corrections to the splitting functions follow from these and are given by
\begin{align}
   \label{eq:pppgl}
    \text{Split}^{(1)}_{\text{fact}} \left(1^{+}, 2^{+}; -P^{+}\right) &= \mA_3^{(1)} \left(1^{+}, 2^{+}, -P^{+}\right) \frac{1}{\an{12}\sq{21}} \\
    \nonumber &=\frac{N_g \,\kappa_{1,1,-1}^3}{16\pi^2}\frac{\sq{P1}\sq{2P}}{\an{12}^2\sq{12}} +\mc{O}(\langle 12\rangle^{-1}).
\end{align}
Consequently, the corresponding splitting function in (\ref{eq:pppgl}) has a double pole at $\an{12} = 0$. This double pole can be predicted from dimensional analysis and little group scaling. The lack of a dimensionful parameter means that the splitting function cannot be proportional to the 3-point amplitude $\mA_3\left(1^{+}, 2^{+}, 3^{+} \right)$, whose coupling $\kappa_{1,1,1}$ has mass dimension $-2$. The only way to construct an expression consistent with little group scaling and with the correct dimension is to introduce an extra pole. Since $\lambda_1 \propto \lambda_2 \propto \lambda_P$, this leaves us with a unique choice. We will use this reasoning to identify the splitting functions that can contain double poles in Section \ref{sec:dpcancel}. Finally, the non-factorizing contributions in (\ref{eq:1loopsplit}) arise from IR divergences in the one-loop amplitude. We will not elaborate on them here other than to say that they give rise to logarithmic terms in the splitting function and refer the reader to \cite{Bern:1995ix} for more details.

\section{Double pole cancellation in supersymmetric theories}
\label{sec:dpcancel}

In this section we demonstrate the cancellation of double poles in the splitting functions of two bosons in $\mc{N}=1$ SUGRA and renormalizable $\mc{N}\ge 1$ global SUSY theories around stable vacua. We focus only on massless loops since double poles can only arise from these. We note that in the SUGRA case the massless 3-point amplitude couplings have mass dimensions $[\kappa_{J_1, J_2, J_3}] \in \{-1,0\}$, and in the global SUSY case they all have $[\kappa_{J_1, J_2, J_3}] = 0$. There are six classes of splitting functions: Split$\left(t_1^{J_1}, t_2^{J_2}; t_P^{J_P}\right)$, Split$\left(t_1^{J_1}, t_2^{J_2}; b_P^{J_P}\right)$, Split$\left(b_1^{J_1}, t_2^{J_2}; t_P^{J_P}\right)$ and the three others obtained by the replacement $t_i^{J_i} \leftrightarrow b_i^{J_i}$.  We will focus on the first three classes as all arguments made for these can be applied {\it mutatis mutandis} to the remaining three. The SWIs force Split$\left(t_1^{J_1}, t_2^{J_2}; t_P^{J_P}\right)$ to vanish. To see this, we start from the SWI
\begin{equation}
    \label{eq:SWI}
    \mA_n \left(\left\lbrace t_1^{J_1}, \dots, t_{n-1}^{J_{n-1}}\right\rbrace, \left\lbrace b_n^{J_n} \right\rbrace\right) = 0.
\end{equation}
Taking the collinear limit $\an{12} \to 0$ of this equation, we get
\begin{multline}
    \sum\text{Split}\left(t_1^{J_1}, t_2^{J_2}; b_P^{-J_P}\right) \mA_{n-1} \left(\left\lbrace t_P^{J_P}, \dots, t_{n-1}^{J_{n-1}}\right\rbrace, \left\lbrace b_n^{J_n} \right\rbrace\right)\\
    +\sum\text{Split}\left(t_1^{J_1}, t_2^{J_2}; t_P^{-J_P}\right) \mA_{n-1} \left(\left\lbrace t_3^{J_3}, \dots, t_{n-1}^{J_{n-1}}\right\rbrace, \left\lbrace b_P^{J_P}, b_n^{J_n} \right\rbrace\right) = 0,
\end{multline}
where the sum is over all possible supermultiplets that can be exchanged. The amplitudes on the first line are zero by (\ref{eq:SWI}) while the amplitudes on the second line are generically nonzero. It follows that
\begin{align}
    \sum\text{Split}\left(t_1^{J_1}, t_2^{J_2}; t_P^{-J_P}\right) \mA_{n-1} \left(\left\lbrace t_3^{J_3}, \dots, t_{n-1}^{J_{n-1}}\right\rbrace, \left\lbrace b_P^{J_P}, b_n^{J_n} \right\rbrace\right) = 0.
\end{align}
Since the splitting functions are universal and the above equation must be true in any theory with $\mathcal{N}=1$ SUSY, we can conclude that the splitting function for each multiplet must individually vanish yielding the desired result,
\begin{equation}
    \text{Split}\left(t_1^{J_1}, t_2^{J_2}; t_P^{J_P}\right) = \text{Split}\left(b_1^{J_1}, b_2^{J_2}; b_P^{J_P}\right) = 0.
\end{equation}
We are left with Split$\left(t_1^{J_1}, t_2^{J_2}; b_P^{J_P} \right)$ and Split$\left(b_1^{J_1}, t_2^{J_2}; t_P^{J_P} \right)$. We first focus on the former. It is useful to first identify the subset of these which can manifest double poles. To this end, we note the following constraints on the splitting function:
\begin{enumerate}
    \item Under a little group transformation, the splitting function behaves in the same way as the corresponding 3-point amplitude.
    \item It must have mass dimension $-1$.
    \item Since all holomorphic spinors are proportional in the collinear limit, as in (\ref{eq:collinearspinorrels}), at leading order all angle brackets can be written in terms of $\an{t_1t_2}$ without loss of generality.
\end{enumerate}
These constraints fix the form of the splitting function to
\begin{equation}
    \label{eq:splitansatz}
    \text{Split}^{(1)} \left(t_1^{J_1}, t_2^{J_2}; b_P^{J_P} \right) = N K \,\frac{\sq{t_1 t_2}^{J_1+J_2-J_P}\sq{t_1 b_P}^{J_2+J_P-J_1}\sq{b_P t_1}^{J_P+J_1-J_2}}{\an{t_1t_2}\sq{t_2 t_1}}
\end{equation}
where $K$ is the product of all couplings involved in the diagram underlying the splitting function and $N$ is an order one number times a power of $\an{t_1t_2}\sq{t_2 t_1}$. Both $N$ and $K$ are little group neutral. Their combined mass dimension must be $\left[N K\right] = 1-J_1-J_2-J_P$ ensuring that the splitting function has dimension $-1$. Note also that $[K] \le 0$ since the couplings are either marginal or irrelevant. A double pole in $\an{t_1t_2}$ might arise if $[NK] < 0$, or equivalently
\begin{equation}
\label{eq:nongravdpcondition}
    J_1+J_2+J_P>1.
\end{equation}
A glance at the list of supermultiplets in (\ref{eq:multiplets}) shows that the only configurations with two top states and one bottom state that satisfy (\ref{eq:nongravdpcondition}) are
\begin{align}
    \label{eq:badconfigs}
    \left(J_1, J_2, J_P\right) = &\left(1,1,0\right), \quad \left(2,0,0\right), \quad \left(2,1,0\right), \quad \left(2,2,0\right),\\
    &\left(2,2,-2\right), \quad \left(2,2,-1\right), \quad \left(2,1,-1\right).\nonumber
\end{align}
For non-gravitational amplitudes $[K]$ is always zero since there are only dimensionless couplings. For gravitational amplitudes, it depends on the number of gravitons in the splitting function under consideration and also on the particle(s) circulating in the loop. We expect double poles only if $[N]<0$. A list of the most singular behaviours as $\an{t_1t_2} \to 0$ of $NK$ for the triplets in (\ref{eq:badconfigs}) is shown in Table \ref{tab:badconfigs}. In some of the cases there is more than one possible loop which gives the same leading singular behaviour and in each line in the table we have only indicated one such contribution.

\begin{table}[htb!]
\centering
     \begin{tabular}{|L|L|}
      \hline
        \left(J_1, J_2, J_P\right) & N  K  \text{ (up to a numerical constant)} \\
        \hline
        \left(1, 1, 0\right) & \kappa_{2,1,-1}\, \kappa_{1,0,0}\, \kappa_{-2,1,0} \\
        \hline
        \left(2, 0, 0\right) & \kappa_{2,1,-1}\, \kappa_{1,0,0}\, \kappa_{-1,0,0}  \\
        \hline
        \left(2, 1, 0\right) & \kappa_{2,-1,0}\, \kappa_{1,1,-1}\, \kappa_{1,0,0} \frac{1}{\an{t_1 t_2}\sq{t_2t1_1}} \\
        \hline
        \left(2,2, 0\right) & \kappa_{2,1,-1}\, \kappa_{2,-1,0}\, \kappa_{0,0,1}\frac{1}{\an{t_1t_2}\sq{t_2t_1}}   \\
        \hline
        \left(2, 2, -2\right) & \kappa_{2,2,-2}^3 \an{t_1t_2}\sq{t_2t_1}  \\
        \hline
        \left(2, 2, -1\right) & \kappa_{2,0,-1}^2\, \kappa_{1,1,-1}\frac{1}{\an{t_1t_2}\sq{t_2t_1}}  \\
        \hline
        \left(2, 1, -1\right) & \kappa_{2,1,-1}\kappa_{1,-1,-1}\kappa_{1,-1,1} \\
        \hline
\end{tabular}
     \caption{A list of the most singular behaviours of one-loop splitting functions.}
     \label{tab:badconfigs}
\end{table}
\begin{figure}
\centering
\begin{tikzpicture}
\begin{feynman}
 \vertex (a) at (0, 0.166667);
 \vertex (b) at (0, 1.166667);
 \vertex (c) at (-0.5, 2.03269) {\(t_2^{+2}\)};
 \vertex (d) at (0.866025, 0.666667);
 \vertex (e) at (0, -0.166667);
 \vertex (f) at (0, -1.16667);
 \vertex (g) at (-0.5, -2.03269) {\(t_1^{+2}\)};
 \vertex (h) at (0.866025, -0.666667);
 \vertex (i) at (2.02073, 0);
 \vertex (j) at (1.1547, 0.5);
 \vertex (k) at (1.1547, -0.5);
 \vertex (l) at (3.02073, 0) {\(b_P^{0}\)};
\diagram*{
 (a) -- [photon, edge label=\(t_{\ell_1}^{+1}\)] (b),
 (b) -- [photon, line width=0.05cm] (c),
 (b) -- [photon, white] (c),
 (b) -- [photon, edge label=\(b_{\ell_2}^{-1}\)] (d),
 (e) -- [photon, edge label'=\(b_{\ell_1}^{-1}\)] (f),
 (f) -- [photon, line width=0.05cm] (g),
 (f) -- [photon, white] (g),
 (f) -- [edge label'=\(t_{\ell_3}^{0}\)] (h),
 (i) -- [photon, edge label'=\(t_{\ell_2}^{+1}\)] (j),
 (i) -- [edge label=\(b_{\ell_3}^{0}\)] (k),
 (i) -- (l),
};
\end{feynman}
\end{tikzpicture}
$\qquad\qquad$
\begin{tikzpicture}
\begin{feynman}
 \vertex (a) at (0, 0.166667);
 \vertex (b) at (0, 1.166667);
 \vertex (c) at (-0.5, 2.03269) {\(t_2^{+2}\)};
 \vertex (d) at (0.866025, 0.666667);
 \vertex (e) at (0, -0.166667);
 \vertex (f) at (0, -1.16667);
 \vertex (g) at (-0.5, -2.03269) {\(t_1^{+2}\)};
 \vertex (h) at (0.866025, -0.666667);
 \vertex (i) at (2.02073, 0);
 \vertex (j) at (1.1547, 0.5);
 \vertex (k) at (1.1547, -0.5);
 \vertex (l) at (3.02073, 0) {\(b_P^{0}\)};
\diagram*{
 (a) -- [photon, edge label=\(t_{\ell_1}^{+1}\)] (b),
 (b) -- [photon, line width=0.05cm] (c),
 (b) -- [photon, white] (c),
 (b) -- [photon, edge label=\(b_{\ell_2}^{-1}\)] (d),
 (e) -- [photon, edge label'=\(b_{\ell_1}^{-1}\)] (f),
 (f) -- [photon, line width=0.05cm] (g),
 (f) -- [photon, white] (g),
 (f) -- [edge label'=\(b_{\ell_3}^{0}\)] (h),
 (i) -- [photon, edge label'=\(t_{\ell_2}^{+1}\)] (j),
 (i) -- [edge label=\(t_{\ell_3}^{0}\)] (k),
 (i) -- (l),
};
\end{feynman}
\end{tikzpicture}
    \caption{The two possible configurations of 3-point amplitudes involved in the construction of the loop integrand for $\mA_3^{(1)}\left(t_1^{+2}, t_2^{+2}, b_P^{0}\right)$.}
    \label{fig:220}
\end{figure}
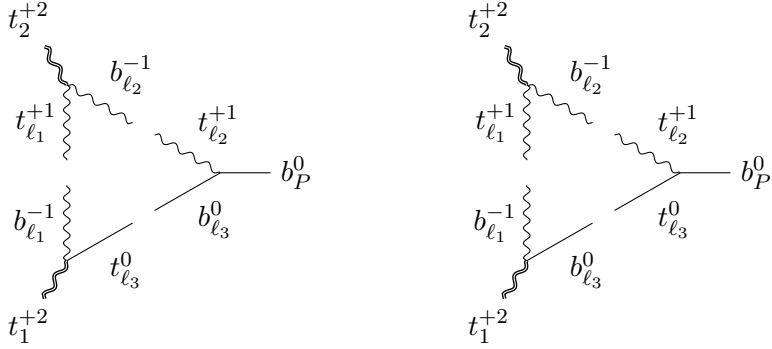

The only potentially troublesome configurations are $\left(J_1, J_2, J_P\right) = \left(2,2,0\right), \left(2,1,0\right),\\ \left(2,2,-1\right)$. However, these cannot be constructed using the tree-level 3-point couplings in $\mathcal{N}=1$ SUGRA. To see this, first consider the $\left(2,2,0\right)$ configuration. Note that the two possible ways of building the integrand from triple cuts are shown in Fig.~\ref{fig:220}.\footnote{It is also possible to construct this amplitude with a scalar in the loop. However, this would require a $\phi^3$ type interaction which we discount on the basis of vacuum stability.} The construction on the left requires $\mA_3^{(0)}(t_{\ell_2}^{+1}, b_{\ell_3}^{0}, b_{P}^{0})$ while the one on the right requires $\mA_3^{(0)}(t_1^{+2}, b_{\ell_1}^{-1}, b_{\ell_3}^{0})$. However, as observed in (\ref{eq:vanishingamps}), these 3-point amplitudes are absent in $\mc{N}=1$ SUGRA. Constructing the integrand from double cuts runs into a similar issue since the 4-point tree-level amplitude required for this construction must in turn be constructed from the 3-point ones shown in Fig.~\ref{fig:220}. A nearly identical argument shows that double poles cannot arise in the other two configurations thus implying that the class Split$\left(t_1^{J_1}, t_2^{J_2}; b_P^{J_P}\right)$ does not contain double poles. Since ($\ref{eq:nongravdpcondition}$) is symmetric in $J_1, J_2, J_P$, most of the arguments presented above also apply to Split$\left(b_1, t_2; t_P\right)$. The configurations which potentially gave rise to double pole are necessarily of the form $\left(t_1, t_2, b_P\right)$. This completes the proof that double poles cannot arise at one-loop in supersymmetric theories. It also proves that double poles are absent in theories with global SUSY and only renormalizable 3-point interactions. We expect such cancellations to hold at all loop orders but leave a thorough investigation of this to the future.

\acknowledgments

We thank Nima Arkani-Hamed, Zvi Bern, Roland Bittleston, Daniel Kapec, Sebastian Mizera, Miguel Montero, Andrzej Pokraka, Lecheng Ren, Oliver Schlotterer, Christian Schubert, and Andrew Strominger for useful discussions. This work was supported in part by the US Department of Energy under contract DE\-SC0010010 (Task F), by Simons Investigator Award \#376208 (AV), by Bershadsky Distinguished Visiting Fellowships at Harvard (MS and AV), and by the STFC grant ST/X000761/1 (AYS).

\appendix

\section{\texorpdfstring{Computation of $\mA_3^{(1),sc}$}{Computation of mA31sc}}
\label{app:3pt-1loop-sc}

\begin{figure}
\centering
\begin{tikzpicture}
\begin{feynman}
 \vertex (a) at (0, 0.166667);
 \vertex (b) at (0, 1.166667);
 \vertex (c) at (-0.5, 2.03269) {\(2^+\)};
 \vertex (d) at (0.866025, 0.666667);
 \vertex (e) at (0, -0.166667);
 \vertex (f) at (0, -1.16667);
 \vertex (g) at (-0.5, -2.03269) {\(1^+\)};
 \vertex (h) at (0.866025, -0.666667);
 \vertex (i) at (2.22073, 0);
 \vertex (j) at (1.3547, 0.5);
 \vertex (k) at (1.3547, -0.5);
 \vertex (l) at (3.24073, 0) {\(3^+\)};
 \vertex (x) at (1.11036, 1.106667);
 \vertex (y) at (1.11036, -1.106667);
\diagram*{
 (b) -- [gluon] (c),
 (b) -- [momentum=\(\ell\)] (d),
 (f) -- [momentum=\(\ell+p_2\)] (b),
 (f) -- [gluon] (g),
 (f) -- [momentum'=\(p_3-\ell\)] (h),
 (i) -- [momentum'=\(-\ell\)] (j),
 (i) -- [momentum=\(\ell-p_3\)] (k),
 (i) -- [gluon] (l),
 (x) -- [scalar] (y),
};
\end{feynman}
\end{tikzpicture}
    \caption{A unitarity cut of $\mA_3^{(1),sc}$.}
    \label{fig:unitaritycuts}
\end{figure}
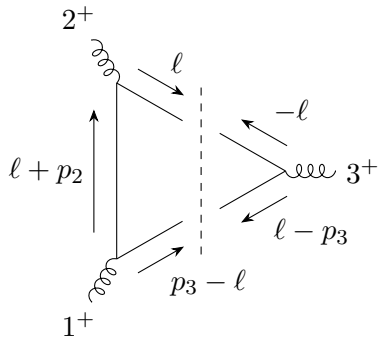
In this appendix, we will compute the one-loop correction to the splitting function of two positive helicity gluons from a massive scalar loop. This serves to demonstrate the fact that massive loops do not give rise to double poles while also setting up some of the  machinery necessary to compute the correction from a gluon loop. The one-loop correction arises from the amplitude $\mA^{(1), sc}_3\left(1^{+}, 2^{+}, \bsym{3}\right)$. Here particles 1 and 2 are positive helicity gluons and particle 3 is an off-shell gluon but we have chosen to treat it as a massive particle instead. We will adopt the convention of labelling massive or off-shell particles in boldface. The particle circulating in the loop is a massive scalar. We are interested in the behaviour of the loop corrections as $p_3^2 \to 0$ and thus need to keep it off-shell to begin with. We first construct the integrand by gluing tree-level amplitudes on a unitarity cut \cite{Bern:1994zx, Anastasiou:2006jv} (see Fig.~\ref{fig:unitaritycuts}):
\begin{multline}
     \mA^{(1), sc}_3\left(1^{+}, 2^{+}, \bsym{3}\right) = \int \frac{d^D \ell}{(2\pi)^D} \mA_4^{(0)} \left(1^{+}, 2^{+}, -\ell, \bsym{p_3}-\ell \right) \mA_3^{(0)} \left( -\ell, \bsym{-p_3}+\ell, \bsym{p_3}\right)\\
     \times\frac{i}{\left(\ell+p_2\right)^2 - m^2}\frac{i}{\left(\ell-p_3\right)^2 - m^2}\frac{i}{\ell^2 - m^2}.
\end{multline}
We can decompose the the $D$ dimensional vector $\ell = \tell + \mu$, with $\tell$ living in the 4D subspace of physical momenta and $\mu$ in its orthogonal space and re-write the above equation as
\begin{multline}
     \mA^{(1), sc}_3\left(1^{+}, 2^{+}, \bsym{3}\right) = \frac{-i}{2\,\Gamma\left(\frac{D-4}{2}\right)(4\pi)^{\frac{D-4}{2}}}\int d\mu^2 \left(\mu^2\right)^{\frac{D-6}{2}}
     \int \frac{d^4 \tell}{(2\pi)^4} \\
     \times\sum_{\text{states}}\frac{\mA_4^{(0)} \left(1^{+}, 2^{+}, -\bsym{\ell}, \bsym{p_3-\ell} \right)\mA_3^{(0)} \left( -\bsym{\ell}, \bsym{-p_3+\ell}, \bsym{p_3}\right)}{((\tell+p_2)^2 - m^2 - \mu^2)((\tell-p_3)^2 - m^2 - \mu^2)(\tell^2 - m^2 - \mu^2)}.
\end{multline}
The sum is over all the states that circulate in the loop. In this case, it is merely a trace over the representation of the scalar. In arriving at this equation, we have utilized the spherical symmetry of the integrand (since all the physical momenta are orthogonal to $\mu$) and performed the angular integral over the $(D-4)$ sphere. Furthermore, we are treating the amplitude as involving massive particles rather than $D$-dimensional ones. We can now simplify this expression by noting that in the amplitudes, we have
\begin{equation}
    \ell^2 = \left(\ell - p_3 \right)^2 = \mu^2 \quad \implies \quad 2\ell \cdot p_3 = p_3^2.
\end{equation}
Consequently, we have the following leading behaviour
\begin{equation}
    \mA_3^{(0)} \left( -\bsym{\ell}^0, \left(\bsym{-p_3+\ell}\right)^0, \bsym{p_3}^1\right) \xrightarrow[]{p_3^2 \to 0} \mA_3^{(0)} \left( -\bsym{\ell}^0, \left(\bsym{-p_3+\ell}\right)^0, p_3^{\pm}\right) + \mo \left(p_3^2\right).
\end{equation}
We can thus use the minimal coupling \cite{Arkani-Hamed:2017jhn} 
\begin{align}
    \mA_3 \left(\bsym{1}^0, \bsym{2}^0, 3^{\pm}\right) =\kappa_{1,0,0} \, \e_3 \cdot p_1
\end{align}
instead of an all-massive 3-point amplitude. Here $\e_3$ is the polarization vector of the gluon. Finally, we also need the 4-point amplitude \cite{Ballav:2020ese, Bern:1995db}
\begin{align}
\label{eq:4pt2g2msc}
    \mA_4^{(0)} \left(1^+, 2^+, \bsym{\ell}^0, \left(\bsym{-\ell+p_3}\right)^0\right) = \kappa_{1,0,0}^2 \frac{\sq{12}}{\an{12}} \frac{m^2+\mu^2}{(p_2+\tell)^2-m^2-\mu^2}.
\end{align}
Putting this together gives
\begin{multline}
\label{eq:1loopmassivesc3pt}
    \mA^{(1), sc}_3\left(1^{+}, 2^{+}, \bsym{3}\right) = -\frac{i\,N_{sc}\,\kappa_{1,0,0}^3}{2\,\Gamma\left(\frac{D-4}{2}\right)(4\pi)^{\frac{D-4}{2}}} \frac{\sq{12}}{\an{12}}\int d\mu^2 \left(\mu^2\right)^{\frac{D-6}{2}} \int \frac{d^4 \tell}{(2\pi)^4}\\
    \times \frac{\left(\mu^2+m^2\right) \,\e_3 \cdot \tell }{(\tell+p_2)^2 - m^2-\mu^2)((\tell-p_3)^2 - m^2-\mu^2)(\tell^2 - m^2-\mu^2)}
\end{multline}
where $\e_3$ is the polarization vector of the off-shell gluon. The integral can be simplified by Feynman parametrization to
\begin{align}
    \nonumber &\mA^{(1), sc}_3\left(1^{+}, 2^{+}, \bsym{3}\right) = -\frac{N_{sc} \,\kappa_{1,0,0}^3 \,\e_3 \cdot p_2}{2\,\Gamma\left(\frac{D}{2}\right)(4\pi)^{\frac{D}{2}}} \frac{\sq{12}}{\an{12}} \int_0^{1} d\alpha_1 \, \int_0^{1-\alpha_1} d\alpha_3\\
    &\qquad\qquad\qquad\qquad\qquad\qquad\times\int d\mu^2 \left(\mu^2\right)^{\frac{D-6}{2}} \frac{\left(\mu^2+m^2\right) \alpha_1 }{\left(p_3^2 \, \alpha_3 \left(1-\alpha_1-\alpha_3\right)-m^2-\mu^2\right)}\\
    &\nonumber \qquad\qquad\qquad= \frac{N_{sc} \, \kappa_{1,0,0}^3 \,\e_3 \cdot p_2}{2\,(4\pi)^{\frac{D}{2}}} \frac{\sq{12}}{\an{12}} \Gamma\left(\textstyle{\frac{6-D}{2}}\right)p_3^2
    \int_0^{1} \, \int_0^{1-\alpha_1} \frac{d\alpha_1 \,d\alpha_3 \,\alpha_3 \left(1-\alpha_1-\alpha_3\right)}{\left(m^2-p_3^2 \, \alpha_3 \left(1-\alpha_1-\alpha_3\right)\right)^{\frac{6-D}{2}}}.
\end{align}
Setting $D = 4-2\varepsilon$ and $p_3^2 =0$ in the integral, we get the leading behaviour
\begin{align}
\label{eq:3pt1loopsc}
    \mA^{(1), sc}_3\left(1^{+}, 2^{+}, \bsym{3}\right) &\xrightarrow[D=4]{p_3^2 \to 0} \frac{N_{sc} \,\kappa_{1,0,0}^3 \,\e_3 \cdot p_2}{48\,(4\pi)^{2}} \frac{\sq{12}}{\an{12}} \frac{p_3^2}{m^2}.
\end{align}
We can now extract the two separate helicity amplitudes by using the polarization vectors
\begin{equation}
    \label{eq:polvecs}
\e_3^+ = \frac{\tilde{\lambda}_3 \, \xi}{\an{3\xi}}, \qquad \e_3^{-} = \frac{\lambda_3 \, \tilde{\xi}}{\sq{3\xi}} .
\end{equation}
which gives
\begin{align}
     &\mA^{(1), sc}_3\left(1^{+}, 2^{+}, 3^{+}\right) \xrightarrow[]{p_3^2 \to 0} \frac{N_{sc} \, \kappa_{1,0,0}^3}{24\,(4\pi)^{2}}  \frac{\sq{12}}{\an{12}} \frac{\an{12}\sq{21}}{m^2} \frac{\sq{32}\an{2\xi}}{\an{3\xi}} =\frac{N_{sc} \,\kappa_{1,0,0}^3}{24\,m^2(4\pi)^{2}} \sq{12}\sq{23}\sq{31},\\
     \nonumber &\mA^{(1), sc}_3\left(1^{+}, 2^{+}, 3^-\right) \xrightarrow[]{p_3^2 \to 0} \frac{N_{sc} \, \kappa_{1,0,0}^3 \,}{24\,(4\pi)^{2}} \frac{\sq{12}}{\an{12}} \frac{\an{12}\sq{21}}{m^2} \frac{\an{32}\sq{2\xi}}{\sq{3\xi}} = \frac{N_{sc} \,\kappa_{1,0,0}^3 \,}{24\,m^2 (4\pi)^{2}} \frac{\sq{12}^2 \sq{23}}{\sq{31}}\an{12} .
\end{align}
In arriving at the above results, we have made use of (\ref{eq:collinearspinorrels}). As expected, these contributions are not singular as $\an{12} \to 0$ and thus do not lead to double poles in the splitting function.

\section{\texorpdfstring{Computation of $\mA_3^{(1)}$}{Computation of mA31gl}}
\label{app:3pt-1loop-gl}

The computation of the correction from the gluon is very similar to the one in the previous appendix. Once again, we construct the integrand by the method of unitarity cuts. Since we are working in $D$ dimensions, and we were interested in obtaining the loop correction from a massless scalar, we would just have to set $m=0$ in (\ref{eq:1loopmassivesc3pt}). In order to obtain the correction from a massless gluon, we must make the following modifications:
\begin{enumerate}
    \item Use the 4-point amplitude with two gluons and two massive spin-1 particles instead of (\ref{eq:4pt2g2msc}) in the integrand.
    \item Use the appropriate 3-point minimal coupling $\mA^{(0)}_3 \left(\bsym{\ell}_1^1,\bsym{\ell}_2^1, 3^{\pm}\right)$.
    \item Sum over the states (polarizations and colours) of the spin-1 particle.
\end{enumerate}
The required amplitudes are \cite{Ballav:2020ese, Arkani-Hamed:2017jhn}
\begin{align}
    \mA_4^{(0)} \left(1^+, 2^+, \bsym{\ell}_1^1, \bsym{\ell}_2^1\right) &= \kappa_{1,0,0}^2 \frac{\sq{12}}{\an{12}} \frac{\an{\bsym{\ell_1 \ell_2}}^2}{(p_2+\tell_1)^2-\mu^2},\\  \mA_3^{(0)} \left(\bsym{\ell}_1^1,\, \bsym{\ell}_2^1, 3^{\pm}\right) &= \frac{\kappa_{1,1,-1}}{\mu^2}\e_3^{\pm} \cdot \ell_1 \an{\bsym{\ell}_1 \bsym{\ell}_2}^{2}.
\end{align}
Here $\mu$ is the mass of the spin-1 particles and we have $\ell_1^2 = \ell_2^2 = \mu^2$ and the angle brackets involve massive spinor helicity variables \cite{Arkani-Hamed:2017jhn}. The object entering the integrand is the sum
\begin{align}
   \nonumber &\sum_{\text{states}}\mA_4^{(0)} \left(1^+, 2^+, \bsym{\ell}_1^1, \bsym{\ell}_2^1\right) \mA_3^{(0)} \left(\bsym{\bar{\ell}}_1^1,\, \bsym{\bar{\ell}}_2^1, 3^{\pm}\right) \\
     &\qquad\qquad= \frac{N_g \,\kappa_{1,1,-1}^3 \e_3^{\pm} \cdot \ell_1}{(p_2+\tell_1)^2-\mu^2} \frac{\sq{12}}{\an{12}}\frac{1}{\mu^2} \sum_{\text{polarizations}}\an{\bsym{\ell_1 \ell_2}}^2\an{\bsym{\bar{\ell}_1 \bar{\ell}_2}}^2\\
    \nonumber &\qquad\qquad= \frac{N_g\, \kappa_{1,1,-1}^3 \e_3^{\pm} \cdot \ell_1}{(p_2+\tell_1)^2-\mu^2} \frac{\sq{12}}{\an{12}} \, \times 3\mu^2 = 3 N_g \,\mA_4^{(0)} \left(1^+, 2^+, \bsym{\ell}_1^0, \bsym{\ell}_2^0\right) \mA_3^{(0)} \left(\bsym{\bar{\ell}}_1^0,\, \bsym{\bar{\ell}}_2^0, 3^{\pm}\right) .
\end{align}
In the above equation, we have used the notation $\bar{\bsym{\ell}}_1$ to denote the fact that within the polarization sum, this particle has the opposite spin orientation to $\bsym{\ell}_1$. $N_g$ is the dimension of the adjoint representation. In arriving at the first equality on the last line, we have made use of the completeness relation $\sum_{\text{polarizations}}\an{\bsym{\ell_1 \ell_2}}^2\an{\bsym{\bar{\ell}_1 \bar{\ell}_2}}^2 = 3\mu^4$ \cite{Caron-Huot:2021iev} which holds since massive particles have three degrees of freedom. The final equality notes that the integrand for the contribution of a massive spin-1 particle is thrice that of a scalar. Since the particle is actually massless and has two degrees of freedom, we must subtract one scalar degree of freedom. With these observations, we can write
\begin{align}
   \nonumber \mA_3^{(1)} \left(1^+, 2^+, 3^{\pm} \right) &= \frac{N_g \,\kappa_{1,0,0}^3 \,\e_3^{\pm} \cdot p_2}{\,(4\pi)^{\frac{D}{2}}} \frac{\sq{12}}{\an{12}} \Gamma\left(\textstyle{\frac{6-D}{2}}\right)\left(p_3^2\right)^{\frac{D-4}{2}} \\
    &\times(-1)^{\frac{D-4}{2}}\int_0^{1} \, \int_0^{1-\alpha_1} d\alpha_1 \,d\alpha_3 \,\alpha_3^{\frac{D-4}{2}}\left(1-\alpha_1-\alpha_3\right)^{\frac{D-4}{2}} \\
    \nonumber &\xrightarrow[]{D=4}\frac{N_g \,\kappa_{1,0,0}^3 \,\e_3^{\pm} \cdot p_2}{\,(4\pi)^{2}} \frac{\sq{12}}{\an{12}}.
\end{align}
Using the polarization vectors in (\ref{eq:polvecs}), we get
\begin{align}
    &\mA_3^{(1)} \left(1^+, 2^+, 3^{+} \right) =\frac{N_g \, \kappa_{1,0,0}^3}{\,(4\pi)^{2}} \frac{\sq{32}\sq{31}}{\an{12}}, \qquad \mA_3^{(1)} \left(1^+, 2^+, 3^{-} \right) =\frac{N_g \, \kappa_{1,0,0}^3}{\,(4\pi)^{2} } \frac{\sq{12}^3}{\sq{23}\sq{31}}.
\end{align}

\bibliography{main}

\bibliographystyle{JHEP}

\end{document}